\documentclass[aps,prb,twocolumn,showpacs]{revtex4}
\usepackage{amsmath}

\usepackage{graphicx}

\begin{document}

\title{Fully gapped superconductivity in In-doped topological crystalline insulator Pb$_{0.5}$Sn$_{0.5}$Te}

\author{Guan Du$^{1}$, Zengyi Du$^1$, Delong Fang$^1$, Huan Yang$^{1*}$, R. D. Zhong$^2$, J. Schneeloch$^2$, G. D. Gu$^2$, and Hai-Hu Wen$^{1}$}\email{huanyang@nju.edu.cn, hhwen@nju.edu.cn}

\affiliation{$^1$National Laboratory of Solid State Microstructures and Department of Physics, Collaborative Innovation Center of Advanced Microstructures, Nanjing University, Nanjing 210093, China}
\affiliation{$^2$Condensed Matter Physics and Materials Science Department, Brookhaven National Laboratory, Upton, New York 11973, USA}

\begin{abstract}
Superconductors derived from topological insulators and topological crystalline insulators by doping have long been considered to be candidates as topological superconductors. Pb$_{0.5}$Sn$_{0.5}$Te is a topological crystalline insulator with mirror symmetry protected surface states on (001), (011) and (111) oriented surfaces. The superconductor (Pb$_{0.5}$Sn$_{0.5}$)$_{0.7}$In$_{0.3}$Te is induced by In doping in Pb$_{0.5}$Sn$_{0.5}$Te, and is thought to be a topological superconductor. Here we report the first scanning tunneling spectroscopy measurement of the superconducting state as well as the superconducting energy gap in (Pb$_{0.5}$Sn$_{0.5}$)$_{0.7}$In$_{0.3}$Te on a (001)-oriented surface. The spectrum can be well fitted by an anisotropic $s$-wave gap function of $\Delta=0.72+0.18\cos4\theta$ meV using Dynes model. The results show that the quasi-particle density of states seem to be fully gapped without any in-gap states, in contradiction with the expectation of a topological superconductor.
\end{abstract}


\maketitle

Topological superconductors (TSCs) have been predicted as a new phase of matter, and are topologically distinct from conventional superconductors\cite{Qi}. As a manifestation of their peculiar properties, topological superconductors host solid-state realizations of Majorana fermions at their boundaries which obey non-Abelian statistics and have potential value of realizing topological quantum computation\cite{Kitaev}.

Chasing for materializing TSCs has already been a very hot spot. One method to obtain topological superconductivity is constructing topological insulator/conventional superconductor heterostructures\cite{Fu and Kane, Wang, on cuprate, Xu and Jia1, Qiang-Hua Wang, Xu and Jia2}. Topological insulators (TIs) are cousins of TSCs with insulating bulk states and metallic surface states protected by time-reversal symmetry\cite{Hasan}. Topological superconductivity can be induced in TIs by superconducting proximity effect\cite{Fu and Kane}. Superconductors derived from TIs by doping have also been considered as candidates, e.g., Cu$_x$Bi$_2$Se$_3$\cite{Hor and Cava} is predicted to be a TSC theoretically\cite{Fu and Berg}. Point contact measurements have detected zero-bias conductance peaks (ZBCPs)\cite{Sasaki and Ando, Kirzhner and Kanigel} which are interpreted to be signatures of Majorana fermions\cite{Hsieh and Fu CuxBi2Se3, Yamakage and Tanaka}. However the ZBCP is absent in normal metal/superconductor junction with finite barriers\cite{Peng and Chu} and scanning tunneling spectroscopy (STS) measurements\cite{Levy and Stroscio}.

Another interesting system to search TSCs is the superconductor derived from topological crystalline insulator (TCI) by doping. Topological crystalline insulators are topologically nontrivial states of matter in which the topological nature of electronic structures arises from crystal symmetries\cite{Fu TCI}. Examples of TCIs have been found in the IV-VI semiconductor SnTe\cite{Hsieh and Fu, Tanaka and Ando}, and substitutional solid solutions Pb$_{1-z}$Sn$_z$Te\cite{Xu and Hasan, Safaei and Buczko} and Pb$_{1-z}$Sn$_z$Se\cite{Dziawa and Story, Wojek and Tjernberg, Pletikosi¡äc and Valla}. Pb$_{1-z}$Sn$_z$Te undergoes a transition from a trivial insulator to a TCI when $z$ exceeds 0.3\cite{Tanaka and Ando 0.3}. The (001), (011) and (111) oriented surfaces of these TCIs are mirror symmetric about \{011\} mirror planes and have an even number of Dirac cones\cite{Safaei and Buczko mirror symmetry, Liu and Fu, Wang and Bansil}, which is quite different with TIs. Superconductivity has been found in In doped SnTe\cite{Zhong and Gu SnInTe, Balakrishnan and Lees}, and is considered to be topologically nontrivial\cite{Zhong and Gu SnInTe}. ZBCPs have been detected by point contact measurements in Sn$_{1-x}$In$_x$Te and are also interpreted as signatures of nontrivial topological superconductivity\cite{Sasaki and Ando SnInTe}. However, some groups reported later that their data do not support the conclusion of a novel superconductor\cite{He and Li, Saghir and Lees}. Whether the superconductivity in doped TCIs is topologically nontrivial is still unknown. Superconductivity was also found in In doped Pb$_{1-z}$Sn$_z$Te\cite{PbSnInTe1, PbSnInTe2, PbSnInTe3, PbSnInTe4, PbSnInTe5, PbSnInTe6, PbSnInTe7}, and this material is a new possible candidate for TSCs according to former theoretical proposals\cite{Sasaki and Ando SnInTe}. The quality of (Pb$_{1-z}$Sn$_z$)$_{1-x}$In$_x$Te crystals and the superconducting transition temperature has been improved recently\cite{Zhong and Gu PbSnInTe}, and $T_c$ reaches maximum 4.7 K at its optimum doping of $x=0.3$. To verify its topological properties accurately, detecting the gap structure is strongly called for.

STS is a direct probe to detect the local density of states (LDOSs), which can provide key and intrinsic information on the superconducting gap symmetry\cite{STM1,STM2}. In this Rapid Communication, we present the first measurements of the local tunneling spectra on optimal doped (Pb$_{1-z}$Sn$_z$)$_{1-x}$In$_x$Te and results of its superconducting gap structure.


\begin{figure}
\includegraphics[width=9cm]{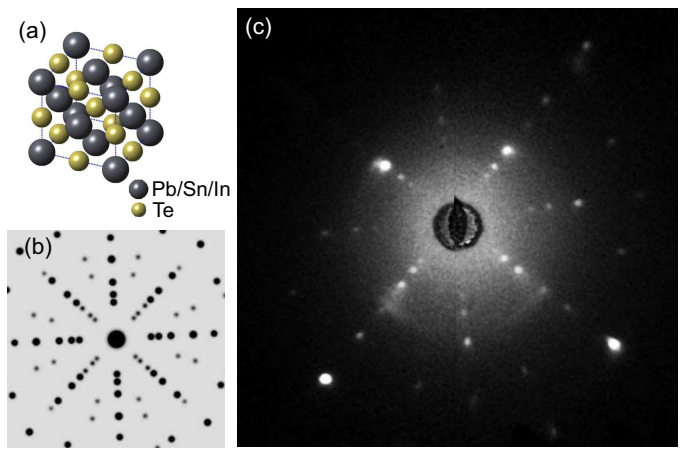}
\caption {(color online) (a) The schematic crystal structure of (Pb$_{0.5}$Sn$_{0.5}$)$_{0.7}$In$_{0.3}$Te. (b) Theoretically simulated Laue diffraction patterns of (001)-oriented surface. (c) Laue diffraction patterns of the cleaved surface of the single crystal.} \label{fig1}
\end{figure}

We chose (Pb$_{0.5}$Sn$_{0.5}$)$_{0.7}$In$_{0.3}$Te for the present investigation, and Sn content of $z = 0.5$ makes it sure that the parent compound is topologically nontrivial as referred above\cite{Tanaka and Ando 0.3}. The high quality single crystal was grown by the floating-zone method\cite{Zhong and Gu PbSnInTe}. The DC magnetic susceptibility was measured using a commercial superconducting quantum interference device magnetometer (SQUID-VSM, Quantum Design). The resistivity was measured using the standard four-probe configuration performed on Physical Property Measurement System (PPMS, Quantum Design).

The crystal structure of (Pb$_{0.5}$Sn$_{0.5}$)$_{0.7}$In$_{0.3}$Te is similar to sodium chloride as shown in Fig.~\ref{fig1}(a). Here Pb, Sn and In atoms share the same kind of sites in the lattice. The (001)-oriented planes are naturally cleavage planes, and are mirror symmetric about \{011\} mirror planes. In parent compound topological crystalline insulator Pb$_{0.5}$Sn$_{0.5}$Te, such kind of surfaces host topologically nontrivial surface states\cite{Safaei and Buczko mirror symmetry}. Therefore identifying the (001) crystal orientation in the first place is very necessary in order to detect underlying features of the superconductivity in (Pb$_{0.5}$Sn$_{0.5}$)$_{0.7}$In$_{0.3}$Te. Lacking of natural crystal surfaces, we polished the crystal and got a smooth surface. Then Laue x-ray crystal alignment system (Photonic Science Ltd.) was used to investigate the orientation of the crystal according to the Laue diffraction patterns. Fig.~\ref{fig1}(b) shows the theoretical simulation of Laue diffraction patterns when the x-ray perpendicular to the (001)-oriented surface, and obviously it has the fourfold symmetry which is consistent with the crystal structure. We adjusted the incidence angle of the x-ray until fourfold symmetric patterns came into view. Then the crystal was polished again according to the former step. After repeating several times, new patterns were obtained as shown in Fig.~\ref{fig1}(c) when the x-ray was applied perpendicular to the top surface. Since they agree well with the theoretical simulation, the obtained top surface is confirmed to be (001)-oriented. Then the (Pb$_{0.5}$Sn$_{0.5}$)$_{0.7}$In$_{0.3}$Te crystal was cleaved in the ultra-high vacuum chamber with pressure better than 10$^{-10}$ torr at $\sim80$ K. The resultant tiny surfaces with the scale of sub-millimeters and different height are usually (001)-oriented. Then the cleaved sample was quickly transferred to the scanning tunneling microscope (STM) head. The STM/STS measurements were carried out with an ultra-high vacuum, low temperature and high magnetic field scanning probe microscope USM-1300 (Unisoku Co., Ltd.). Pb$_{1-z}$Sn$_z$Se crystal seems very easy to cleave and is usually atomically resolved on the cleaved surface\cite{Madhavan1, Madhavan2, Madhavan3}. However for (Pb$_{0.5}$Sn$_{0.5}$)$_{0.7}$In$_{0.3}$Te the cleavage made many tiny (001)-oriented surfaces, and the measured surfaces usually have the roughness of about several angstroms which may result from the indium doping. The STS spectra were measured by a lock-in amplifier with an ac modulation of $0.1\;$mV at $987.5\;$Hz to lower down the noise.


\begin{figure}
\includegraphics[width=9cm]{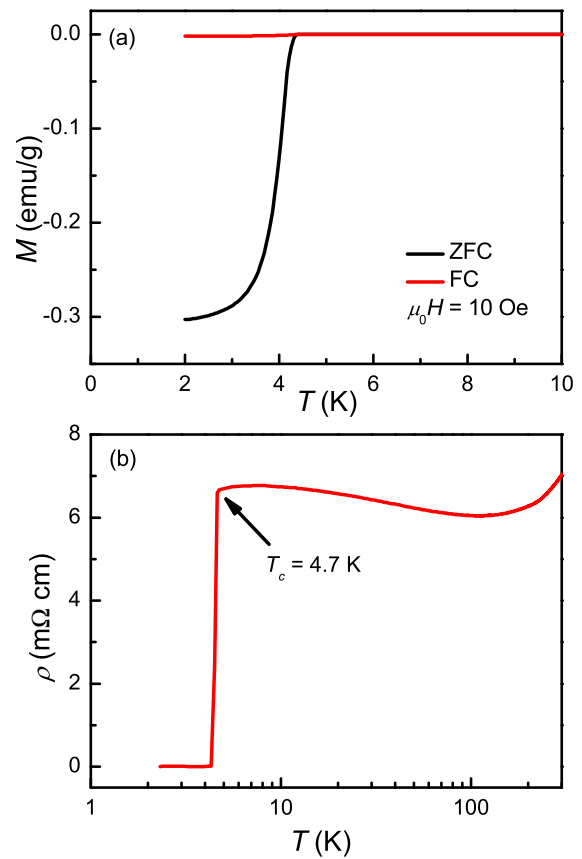}
\caption {(color online) (a) Temperature dependence of magnetic susceptibility for a (Pb$_{0.5}$Sn$_{0.5}$)$_{0.7}$In$_{0.3}$Te single crystal measured with ZFC and FC processes at an applied field of 10 Oe. (b) Temperature dependence of the resistivity in semi-log plot. } \label{fig2}
\end{figure}

Fig.~\ref{fig2}(a) shows the temperature dependence of magnetic susceptibility for a (Pb$_{0.5}$Sn$_{0.5}$)$_{0.7}$In$_{0.3}$Te single crystal and the superconducting transition occurs at about 4.5 K characterized by the zero-field-cooled (ZFC) process and field-cooled (FC) process at 10 Oe. Temperature dependence of the electrical resistivity is plotted in Fig.~\ref{fig2}(b). One can find that temperature dependent resistivity increases with the temperature decreasing from 110 K to 7 K. This semiconductor like behavior is rare for superconductors, and may be originated from the insulating behavior in the parent compound of Pb$_{0.5}$Sn$_{0.5}$Te. The superconducting transition occurs at about 4.7 K with a narrow transition width, indicating the high quality of the sample.

\begin{figure}
\includegraphics[width=9cm]{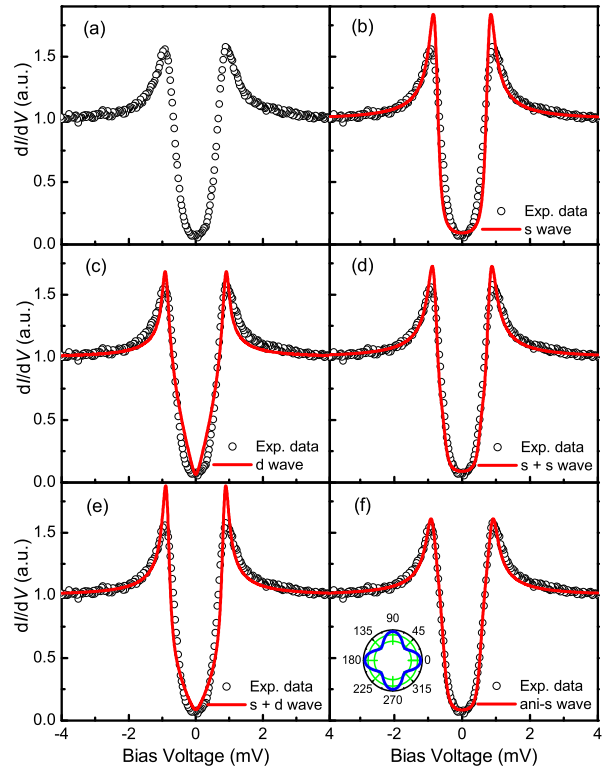}
\caption {(color online) (a) A typical STS spectrum measured at 400 mK. The fitting results are shown in (b)$-$(f). The symbols represent the experimental data, and the colored lines are the theoretical fits to the data with the Dynes model by (b) $s$-wave, (c) $d$-wave, (d) double $s$-waves, (e)$d + s$ wave and (f) an anisotropic $s$-wave gap, respectively. The inset in (f) shows a fourfold symmetric gap function $\Delta=0.72+0.18\cos2\theta$ meV yielded from the fitting with an anisotropic $s$-wave gap.} \label{fig3}
\end{figure}

We applied STS measurements on the newly cleaved (001)-oriented surface at the temperature of 400 mK. A typical spectrum is shown in Fig.~\ref{fig3}(a) covering a voltage range of $\pm4$ mV, and it is  symmetrical to the zero-bias in the presented range. The superconducting gap appears near the Fermi level with pronounced coherence peaks located at $\pm0.9$ mV. The `U'-shape instead of a `V'-shape of the bottom near the zero bias suggests a nodeless superconducting gap existing in the material. The differential conductance value at zero-bias is about 0.07 with relative to the value at higher bias voltage outside coherence peaks, indicating that the LDOSs within the energy of superconducting gaps are almost fully gapped. Odd paring symmetries are predicted theoretically, and the bulk LDOSs can be either fully gapped or gapped with point node/line node(s) by tuning the chemical potential and the effective mass of the energy band\cite{Hsieh and Fu, Yamakage and Tanaka}. The surface Andreev bound states appear as helical Majorana fermions in the bulk quasi-particle gap. This kind of in-gap states has several kinds of structure in the energy dispersions, such as cone, caldera, ridge, or valley shaped\cite{Yamakage and Tanaka}. In any cases, those in-gap states give rise to robust zero bias peaks in LDOSs on the surface. Therefore, the zero bias peaks should be viewed in the tunneling conductance measurements as a signature of the nontrivial superconductivity. However, such a finite LDOSs at zero bias is absent in our measurements.

In order to verify the paring symmetry of (Pb$_{0.5}$Sn$_{0.5}$)$_{0.7}$In$_{0.3}$Te, we fitted the data with several scenarios of superconducting gaps based on the Dynes model as shown in Fig.~\ref{fig3}(b)-(f). The results based on pure $s$-wave and $d$-wave fittings are plotted with the experimental data in Fig.~\ref{fig3}(b) and (c) respectively. The pure $s$-wave fitting to the spectrum yields a superconducting gap value of $\Delta=0.77$ meV and broadening parameter of $\Gamma= 0.07$ meV. And for pure $d$-wave fitting, $\Delta=0.9$ meV, $\Gamma= 0.005$ meV. Both of the two fitting results fail to track the low energy feature, i.e., the $s$-wave fitting displays a more flat bottom near the Fermi energy while the $d$-wave fitting appears to be not flat enough. In addition the coherence peaks of the two fitting results are too sharper than the experimental data. Considering possible multi-band in this compound, we also used two mixed components ($s_1 + s_2$ wave or $s + d$ wave) to fit the experimental data (Fig.~\ref{fig3}(d) and (e)). The double $s$-wave fitting yields a mixture of 80\% $s_1$-wave and 20\% $s_2$-wave, with $\Delta_{s_1}=0.80$ meV, $\Gamma_{s_1}= 0.07$ meV; $\Delta_{s_2}=0.55$ meV, $\Gamma_{s_2}= 0.05$ meV. The $s + d$ wave fitting yields a mixture of 50\% $d$-wave and 50\% $s$-wave, with $\Delta_d=0.80$ meV, $\Gamma_d= 0.03$ meV; $\Delta_s=0.85$ meV, $\Gamma_s= 0.03$ meV. The fittings are better than those by one gap as shown in Fig.~\ref{fig3}(b) and (c), but still not perfect. For the (001)-oriented surface being fourfold symmetric, fitting the spectrum using an anisotropic $s$-wave gap with fourfold symmetry is reasonable. As shown in Fig.~\ref{fig3}(e), the anisotropic $s$-wave model expressed as $\Delta=\Delta_1+\Delta_2\cos4\theta$ can fit the experimental data perfectly and catch the features both of the coherence peaks and the bottom. The best fitting leads to $\Delta_1=0.72$ meV, $\Delta_2=0.18$ meV, and $\Gamma=0.058$ meV. The inset of Fig.~\ref{fig3}(e) shows the resultant cruciate-flower-shaped gap function with a maximum of 0.90 meV and a minimum of 0.54 meV. Taking the average gap value we determine $2\Delta_1/k_\mathrm{B}T_\mathrm{c}\approx3.55$ which is close to 3.53 predicted by Bardeen-Cooper-Schrieffer (BCS) theory in the weak coupling regime.

\begin{figure}
\includegraphics[width=9cm]{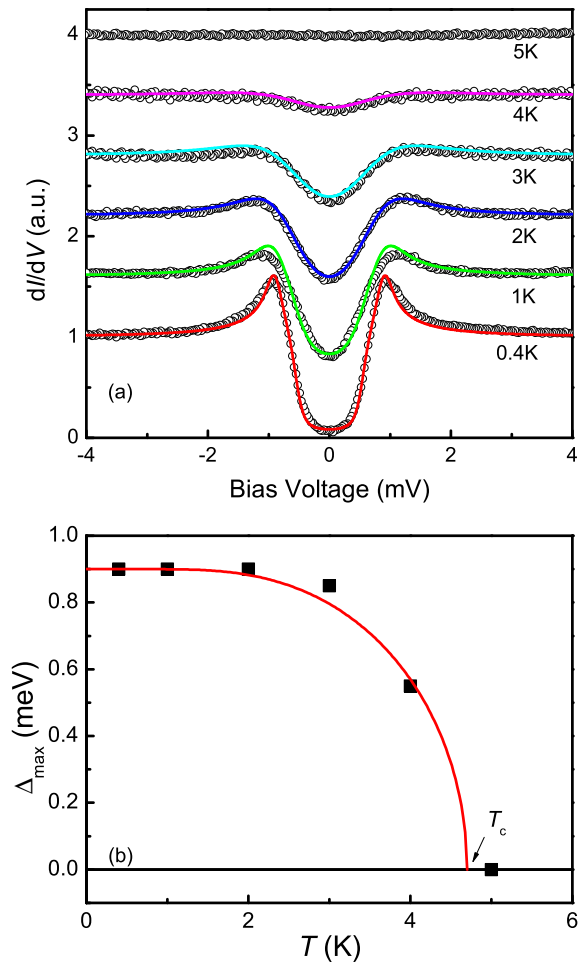}
\caption {(color online) (a) Temperature dependence of tunneling spectra measured from 0.4 K to 5 K (symbols) and theoretical fittings (solid lines) by Dynes model. (b) Temperature dependence of maximal gap value derived by the anisotropic $s$-wave fitting. 
The solid line is the result of theoretical calculation using BCS model by fixing $\Delta_\mathrm{max}(0)$ and $T_\mathrm{c}$ derived from our experiment (see text).} \label{fig4}
\end{figure}

Fig.~\ref{fig4}(a) shows the temperature dependence of tunneling spectra measured from 0.4 K to 5 K as symbols. From a deep groove-like structure, the superconducting feature evolves into a shallow with the increase of the temperature and finally disappears at 5 K which is above $T_\mathrm{c}$. We also fitted the temperature dependent spectra with Dynes model by using anisotropic $s$-wave gaps with constant $\Delta_1/\Delta_2$, and these fittings consist well with the experimental data. The maximum gap values $\Delta_\mathrm{max}=\Delta_1+\Delta_2$ obtained by fitting are plotted in Fig.~\ref{fig4}(b). The solid line in Fig.~\ref{fig4}(b) is obtained through the numerical solution to the BCS gap equation by fixing the $\Delta_\mathrm{max}(0)\approx\Delta_\mathrm{max}(T=0.4\ \mathrm{K})=0.9$ meV and $T_\mathrm{c}=4.7$ K derived from Fig.\ref{fig2}(b). We can find that the temperature dependent energy gap from the fitting to the tunneling spectra can be described well by the BCS model quite well.

We notice that several groups have applied point contact measurements on some candidates of TSCs such as Cu$_x$Bi$_2$Se$_3$ and Sn$_{1-x}$In$_x$Te single crystals. Zero bias conduction peaks were detected\cite{Sasaki and Ando, Kirzhner and Kanigel, Sasaki and Ando SnInTe} and were interpreted in terms of nontrivial topological superconductivity\cite{Hsieh and Fu, Sasaki and Ando SnInTe}. However, due to the complexity of the junctions in point contact experiments, such kind of zero bias peaks can be induced by several mechanisms other than Majorana modes. The STS spectra we measured are fully gapped without any in gap states, which is different from the point contact measurements in other materials mentioned above. Similarly, Levy et al. applied STM experiments on Cu$_x$Bi$_2$Se$_3$ single crystals\cite{Levy and Stroscio}, and a fully gapped superconducting gap was also observed. Another interesting result on Cu$_x$Bi$_2$Se$_3$ shows that the zero bias peaks observed in point contact experiments depend on the barrier transparency and they may not represent the intrinsic feature of the compound\cite{Peng and Chu}. Although Sn$_{1-x}$In$_x$Te is a candidate of TSCs according to theory\cite{Sasaki and Ando SnInTe}, a recent thermal conductivity measurement proves that this compound has a full superconducting gap in the bulk\cite{He and Li}. Another group studied Sn$_{1-x}$In$_x$Te ($x = 0.38\sim0.45$) using muon-spin spectroscopy, and the results can be well described by a single-gap $s$-wave BCS model\cite{Saghir and Lees}. These results have reduced the possibility of being a TCS for Sn$_{1-x}$In$_x$Te. (Pb$_{1-z}$Sn$_z$)$_{1-x}$In$_x$Te and Sn$_{1-x}$In$_x$Te share the same crystal structure, and with our results reported here, we conclude that the superconductivity in (Pb$_{0.5}$Sn$_{0.5}$)$_{0.7}$In$_{0.3}$Te is topologically trivial and even conventional. Since our present study is based on a top layer with the (001) orientation, we cannot exclude the possibility of topological surface state at the surfaces with the (011) and (111) orientations, which will stimulate the effort in the future.


In summary, we report the first set data of temperature dependent scanning tunneling spectroscopy on (Pb$_{0.5}$Sn$_{0.5}$)$_{0.7}$In$_{0.3}$Te, a possible candidate of the topological superconductor. The low temperature spectrum shows the fully-gapped feature without any in-gap states, and it can be fitted well by an anisotropic $s$-wave gap using Dynes model. Combining with previous experiments in Sn$_{1-x}$In$_x$Te with the same structure, we conclude that the (001)-oriented surface we detected in (Pb$_{0.5}$Sn$_{0.5}$)$_{0.7}$In$_{0.3}$Te does not show evidence of topological superconducting state, although the bulk Pb$_{0.5}$Sn$_{0.5}$Te has been proved to be a topological crystalline insulator.

We appreciate the useful discussions with Qianghua Wang and the useful help of using the Laue diffraction machine by Jianzhong Liu. This work was supported by the Ministry of Science and Technology of China (973 projects: 2011CBA00102, 2012CB821403), NSF of China and PAPD. Work at Brookhaven is supported by the Office of Basic Energy Sciences, Division of Materials Sciences and Engineering, U.S. Department of Energy under Contract No. DE-SC00112704. RZ and JS are also supported as part of the Center for Emergent Superconductivity, an Energy Frontier Research Center funded by the U.S. Department of Energy, Office of Science.

\end{document}